\documentclass[fleqn,usenatbib]{mnras}

\usepackage{newtxtext,newtxmath}

\usepackage[T1]{fontenc}
\usepackage{ae,aecompl}

\usepackage{graphicx}	
\usepackage{amsmath}	
\usepackage{amssymb}	

\title[21-cm signal and long-mode modulation]{ 21-cm power spectrum and ionization bias as a probe of long-mode modulated non Gaussian sky}

\author[S. Khosravi et al.]{
Shahram Khosravi,$^{1}$  Amirabbas Ghazizadeh$^{2}$
and Shant Baghram,$^{2}$ \thanks{E-mail: baghram@sharif.edu}
\\
$^{1}$Department of Astronomy and High Energy Physics, Faculty of Physics, Kharazmi University, Tehran, Iran.\\
$^{2}$Department of Physics, Sharif University of Technology, P.~O.~Box 11155-9161, Tehran, Iran.\\
}

\date{Accepted XXX. Received YYY; in original form ZZZ}

\pubyear{2018}

\begin{document}

\def\be{\begin{equation}}
\def\ee{\end{equation}}

\label{firstpage}
\pagerange{\pageref{firstpage}--\pageref{lastpage}}
\maketitle

\begin{abstract}
The observed hemispherical power asymmetry in cosmic microwave background radiation can be explained by long wavelength mode (long-mode) modulation. In this work we study the prospect of the detection of this effect in the angular power spectrum of 21-cm brightness temperature. For this task, we study the effect of the neutral Hydrogen distribution  on the angular power spectrum. This is done by formulating the bias parameter of ionized fraction to the underlying matter distribution. We also discuss the possibility that the long mode modulation is companied with a primordial non-Gaussianity of local type. In this case, we obtain the angular power spectrum with two effects of primordial non-Gaussianity and long mode modulation. Finally, we show that the primordial non-Gaussianity enhances the long mode modulated power of 21-cm signal via the non-Gaussian scale-dependent bias up to four orders of magnitude. {Accordingly, the observation of 21-cm signal with upcoming surveys such as the Square Kilometer Array (SKA) is probably capable of detecting hemispherical power asymmetry in the context of the long mode modulation.}
\end{abstract}




\section{Introduction}
The standard model of cosmology is in good agreement with the observation of the cosmic microwave background radiation (CMB) \cite{Ade2015,Aghanim:2018eyx} and {the results of the surveys of large scale structure (LSS) \cite{Tegmark:2003ud,Eisenstein:2005su,Tegmark:2006az,Anderson:2012sa,Dawson:2015wdb}. Additionally}, the precise data of the CMB from Planck collaboration shows that the initial condition of the Universe is very simple: nearly Gaussian, isotropic, adiabatic and nearly scale invariant \cite{Ade:2013nlj}. This imposes serious constraints on the early universe models including inflationary scenario \cite{Ade:2015lrj,Akrami:2018odb}. Despite the success of the standard model of cosmology, the nature of dark energy, dark matter and the physics of early Universe is still unknown. Also there exist further evidences from CMB sky indicating anomalies such as the cold spot, quadrupole-octupole alignment, power deficit in low multipoles, and {hemispherical power asymmetry} \cite{Ade2016a,Akrami:2019bkn}. These anomalies could be due to the statistical uncertainties, or they can open up a new horizon in the study of the EU beyond the standard picture of $\Lambda$CDM cosmology. In the case of these anomalies pointing toward a new physics,  we can also use the LSS data in the late time Universe to probe the fingerprint of these deviations \cite{Abolhasani:2013vaa, Baghram:2014nha, Namjoo:2014pqa}. However, the LSS data suffers from complex non-linearities, biased observations of baryons, and redshift space distortion (RSD) effect, which must be controlled. Despite these challenges in LSS observation, if we find any specific fingerprint of the EU physics in late time observations, we may be able to probe the deviations from the standard picture of EU inflationary scenario in sub-CMB scales.
In this direction, the galaxies are the first targets of LSS observables to probe EU effects \cite{Zhai:2017ibd}. Another promising area of research is the future 21-cm signal observations which will map the neutral Hydrogen distribution in the universe and also can be used to find the statistics of the distribution of matter \cite{Loeb:2003ya}.
Here, we focus on the distribution of neutral Hydrogen in the epoch of reionization (EoR) to probe the power asymmetry observed on the CMB sky. The idea is that the neutral Hydrogen can be used as a fair sample of matter distribution in large scales \cite{Barkana:2001}. One of the interesting anomalies which is the subject of this work is the CMB {hemispherical power asymmetry}. This asymmetry was first seen in WMAP data \cite{Eriksen:2003db,Eriksen:2007pc,Hansen:2008ym, Hoftuft:2009rq} and it seems that there are indications in Planck analysis as well \cite{Akrami:2014eta,Ade2016a}. In the theoretical side, there are different proposals to describe this asymmetry as long-wavelength mode (long-mode) modulation \cite{Erickcek:2008sm}, domain walls inspired models \cite{Jazayeri:2014nya}, etc. In the present study we have concentrated on the long mode modulation.
If this asymmetry is due to a new physics, and with the assumption that this {feature} will survive in late time (e.g. the feature does not decay with the cosmic time), it is possible, that it affects the late time observables. We assert that the 21-cm cosmology gives the chance to probe the distribution of matter in high redshifts where the matter distribution and evolution can be studied in linear regime. The only issue which must be studied with care is the bias between distribution of neutral Hydrogen and the underlying dark matter. In this direction there are studies that shows the {hemispherical power asymmetry} due to long mode modulation is correlated to the local-type non Gaussianity. In other words, long mode modulation and non-Gaussianity appear simultaneously \cite{Namjoo:2013fka,Namjoo:2014nra}.
Accordingly, we study the effect of long mode modulation on the angular power spectrum of 21-cm brightness temperature with/without non-Gaussian effect on the bias. Then we assert that large scale surveys of 21-cm signal can probe the fingerprint of this long mode on the power spectrum of neutral Hydrogen when local type non-Gaussianity is present. It is worth to mention that the angular power spectrum of 21-cm signal is proposed as a probe of primordial spectral running as well \cite{Sekiguchi:2017cdy}. Also, we should note that \cite{Shiraishi:2016omb} propose to use "off-diagonal components of the angular power spectrum of the 21-cm signal fluctuations during the dark ages to test this power asymmetry". However, in this work we study the angular power spectrum and its correction due to long mode modulation and {Primordial Non-Gaussianity (PNG)}.

The structure of this work is: In Sec. (\ref{sec:longmode}), we briefly introduce the long mode modulation. In Sec. (\ref{sec:21cmpower}) we study the 21-cm power spectrum and scale independent bias. In Sec. (\ref{sec:angular}), we see the effect of the long mode modulation on the angular power spectrum of 21-cm signal. In Sec. (\ref{sec:PNGbias}), the effect of non Gaussianity is studied and finally in  Sec. (\ref{sec:conclusion}) we conclude. In appendix (\ref{sec:modulatedbias}) we study different approximations for bias parameter and its scale dependent case. In appendix (\ref{app2}), we study the characteristics of the SKA experiment and its systematics on the angular power spectrum. Also, the cosmic variance is discussed briefly in this appendix.

In this work we set the present value of matter density parameter $\Omega_m=0.32$, cosmological constant density parameter $\Omega_\Lambda=0.68$, the present value of Hubble parameter $H_{0}=67\text{km}~{\text{s}}^{-1}{\text{Mpc}}^{-1}$ and amplitude of primordial curvature perturbation $A_s=2.1\times10^{-9}$ with curvature perturbation dimensionless power spectrum in the form of ${\cal{P}}_{\cal{R}}(k)=A_{s}{(\frac{k}{k_{0}})}^{n_{s}-1}$ with {spectral index} $n_s=0.965$ and {pivot scale} $k_{0}=0.05\text{Mpc}^{-1}$ \cite{Aghanim:2018eyx}. {The amplitude of modulation in curvature perturbations defined in equation (\ref{eq:curvature_lm}) is set to $A_{\cal{R}}=0.06$ throughout the paper and all figures.}


\section{Theoretical Background}
\label{sec:theoretical}
In this section we set the scene and review the theoretical background. The first subsection is devoted to the long mode modulation, in which we mainly follow the proposal of \cite{Zibin:2015ccn}. In the second subsection we go through the 21-cm brightness temperature fluctuations and we introduce the power spectrum via the bias parameter.
\subsection{Long mode modulation }
\label{sec:longmode}
As mentioned in the introduction, the recent observations of Planck on the {hemispherical power asymmetry} of the CMB, which was first detected by WMAP \cite{Eriksen:2003db}, \cite{Eriksen:2007pc,Hoftuft:2009rq,Hansen:2008ym}, may point toward an anomalous primordial universe. Long mode modulations may be considered as one of the probable explanations of this observation.
The idea of the long mode modulation comes from the {dipole modulation observed} in CMB temperature as \cite{Gordon:2006ag}
\be
\label{dipole-power0}
\Delta T_{\text{CMB}} (\hat{\bf n})=\overline{ \Delta T }_{\text{CMB}}(\hat{\bf n}) \left[1+A_d\cos(\theta_{\hat{\bf n}.\hat{\bf p}}) \right] ,
\ee
in which $\overline{ \Delta T }_{\text{CMB}}(\hat{\bf n}) $ is the statistically isotropic temperature fluctuation,
$A_d$ is the amplitude of the dipolar modulation on temperature template, $ \hat{\bf p}$ is the preferred direction and $\hat {\bf n}$ is the direction of the observation and consequently $\theta_{\hat{\bf n}.\hat{\bf p}}$ is the angle between $ \hat{\bf p}$ and $ \hat{\bf n}$. Planck collaboration has found $A_d \simeq0.06$ and $\hat{\bf p}(l\sim 227^{\circ} ,b \sim -27^{\circ})$ in galactic coordinates \cite{Ade:2013nlj} (For a more detailed analysis see also \cite{Ade2016a,Akrami:2014eta, Aslanyan:2014mqa,Notari:2013iva}).
The idea of long mode modulation, which is the best yet known solution for {hemispherical power asymmetry} \cite{Erickcek:2008sm, Dai:2013kfa}, can produce observational consequences. In this picture, a long super-horizon mode with the wavelength $1/k_L$ modulates the curvature perturbations at the desired scales (either the CMB or the LSS scales, for example in \cite{Abolhasani:2013vaa} the effect of long mode modulation in late time observations is discussed). In this respect, a very important observation is the fact that the modulation dies at large multipoles say $\ell>65$, therefore, it seems reasonable to write the curvature perturbation in large  and small scales as $\tilde{\cal{R}}(x)= \tilde{\cal{R}}^{\text{lo}}(x)+{\cal{R}}^{\text{hi}}(x)$ ("lo" superscript indicating the low wavenumber (i.e. large scales), and "hi" indicating the high multipoles (i.e. small scales)). {One of the possible ways to incorporate this scale dependence is proposed by \cite{Zibin:2015ccn}. In this picture the large scale curvature in real space is modulated as}
\be\label{eq:curvature_lm}
\tilde{{\cal{R}}}^{\text{lo}}(\bmath{x})= {\cal{R}}^{\text{lo}}(\vec{x})\left(1+A_{\cal{R}}\frac{\boldsymbol{r}}{r_{\text{LS}}}.\hat{p}\right),
\ee
where $A_{\cal{R}}$ is the amplitude of modulation in curvature perturbations (i.e. $A_{\cal{R}}=A_{d}$) and $r_{\text{LS}}$ is the distance to the last scattering surface and "$\sim$" indicates modulated quantities. { We should note that there are other phenomenological models which can produce the desired scale dependence. However, in this model there is a reasonable assumption that the modulation dies off in low redshifts.} It is worth to mention that the 21-cm power spectrum is a suitable proposal for study because $r(z\sim 10) / r_{\text{LS}} \sim 1/3$ and it is much more prominent than the low redshift galaxy distribution.
In order to follow this proposal further, we can define the dimensionless unmodulated curvature power spectrum of low and high multipoles from Fourier modes as
\be
\langle {\cal{R}}^{\text{lo}/\text{hi}}(\vec{k}) {\cal{R}}^{\text{lo}/\text{hi}} (\vec{k}')\rangle = \frac{2\pi^2}{k^3}{\cal{P}}^{lo/hi}_{\cal{R}}(k)\delta^{3D}(\vec{k}-\vec{k}'),
\ee
where ${\cal{P}}^{\text{lo}/\text{hi}}_{\cal{R}}(k)$ is the dimensionless unmodulated curvature power spectrum of low and high multipoles. As we mentioned earlier, to explain the suppression of long mode in high angular multipoles, we've divided the curvature perturbation to two parts; also the total statistically isotropic part ${\cal{R}} (k) \equiv {\cal{R}}^{\text{lo}} (k) + {\cal{R}}^{\text{hi}} (k) $ should give the standard $\Lambda$CDM power spectrum as ${\cal{P}}_{\cal{R}}^{\Lambda CDM}({k}) = {\cal{P}}_{\cal{R}}^{\text{lo}} ({k})+ {\cal{P}}_{\cal{R}}^{\text{hi}}({k})$. For this reason the unmodulated large scale curvature perturbation could have the dimensionless power spectra as
\be
{\cal{P}}^{\text{lo}}_{\cal{R}}(k)= \frac{1}{2}{\cal{P}}^{\Lambda CDM}_{\cal{R}}(k)\left[1- \tanh\left(\frac{\ln k - \ln k_{c}}{\Delta \ln k}\right)\right],
\ee
{where $k_c$ is the cut-off wavenumber. Again we should note that the $\tanh$ is just a proposed ansatz to relate the high and low multipoles of power spectrum, other proposals can be made. In order to obtain a modulation on angular scales larger than $l \sim 65$
 one should take $k_c\simeq 5\times 10^{-3} Mpc^{-1}$ as $\chi_* \times k_c = l$ (where $\chi_*$ is the comoving distance to last scattering surface) and $\Delta \ln k \rightarrow 0$,\footnote{In this work we set $\Delta lnk=0.1$ to produce the corresponding plots. This specific choice only affects the smoothing scale of modulated power spectrum.} \cite{Zibin:2015ccn}. }
 The Fourier transform of low multipole curvature is $\tilde{{\cal{R}}}^{\text{lo}}(\vec{k})= {\cal{R}}^{\text{lo}}(\vec{k})+\int d^{3}x{\cal{R}}^{\text{lo}}(\vec{x})A_{\cal{R}}\frac{1}{r_{\text{LS}}}\frac{-1}{i}\frac{\partial}{\partial k_z} exp(-i\vec{k}.\vec{x})$, where the $\hat{z}$ direction is parallel to asymmetry direction $\hat{p}$ so $\tilde{{\cal{R}}}^{\text{lo}}(\vec{k}) = {\cal{R}}^{\text{lo}}(\vec{k}) + i \frac{A_{\cal{R}}}{r_{LS}} \frac{\partial}{\partial k_z} {\cal{R}}^{\text{lo}}(\vec{k})$. { Thus the total power to first order of $A_{\cal{R}}$ can be obtained as \footnote{Note that we take ${\cal{R}}^{\text{lo}}$ and ${\cal{R}}^{\text{hi}}$ to be uncorrelated.}
\begin{equation}
\begin{aligned}
\langle {\tilde{\cal{R}}}(\vec{k}) \tilde{\cal{R}} (\vec{k}')\rangle &= \frac{2\pi^2}{k^3}{\cal{P}}^{\Lambda CDM}_{\cal{R}}(k)\delta^{3D}(\vec{k}-\vec{k}') + 2\pi^2 i \frac{A_{\cal{R}}}{r_{\text{LS}}}  \\
&\times  \left[\frac{{\cal{P}}^{\text{lo}}_{\cal{R}}(k)}{k^3} + \frac{{\cal{P}}^{\text{lo}}_{\cal{R}}(k')}{k'^3}\right] \delta^{2D}(\vec{k}_{\perp} - \vec{k'}_{\perp}) \delta^{\prime}(\vec{k}_{z} - \vec{k'}_{z}),
\end{aligned}
\end{equation}
where $\vec{k}_{\perp}$ is the wavenumber perpendicular to the $z$-direction and prime on Dirac delta function shows the derivative with respect to the argument.}
This is essential to relate the 21-cm power spectrum to the primordial curvature power, which will be discussed in the following sections.
\subsection{21 cm power spectrum and bias}
\label{sec:21cmpower}
One of the promising {observational probes} to trace the distribution of the matter is the map of 21-cm brightness temperature. PNG and long mode modulation can affect the distribution of the early star forming regions. Another advantage of 21-cm signal maps is that we can explore the matter distribution in larger scales.
Accordingly, in this section we will explore the theoretical background of 21-cm signal statistics.

 Here we will focus on the study of the brightness maps at EoR  where the spin temperature $T_s$ is much larger than the CMB temperature ($T_s\gg T_{\text{CMB}}$).
The brightness temperature signal of intergalactic medium (IGM) at redshift $z$  in EoR is \cite{Field:1958,Madau:1996cs}
\be
\Delta T_b \simeq (28 \text{mK}) \left(\frac{\Omega_b h^2}{0.02}\right) \sqrt{\frac{0.15}{\Omega_m h^2}\frac{1+z}{10}}x_{\text{HI}}(1+\delta_{b}),
\ee
where $\Omega_b$ and $\Omega_m$ are baryon and total matter density parameters, $x_{\text{HI}}$ is neutral fraction of baryons in IGM and $\delta_{b}$ is density contrast of baryons at IGM. The
 21-cm fluctuations from IGM at line of sight direction $\hat{n}$  in Fourier space ($\delta T_b(\boldsymbol{k},\hat{n}, z)=\Delta T_b(\boldsymbol{k},\hat{n}, z)-\bar{\Delta T_b}(z)$) is written as
\begin{equation} \label{eq:delta_t_1}
\delta T_b = \bar{\delta T_b}\bar{x}_{\text{HI}}(\delta\rho _{\text{HI}} +  \delta\rho_{m}\mu^2_k),
\end{equation}
where $\delta\rho _{\text{HI}}$ is neutral Hydrogen density fluctuations, $\delta\rho_{m}$ is total matter density fluctuations and $\mu_k\equiv \vec{k}.\hat{n} / |\vec{k}|$ ($\hat{n}$ is the unit vector along the line of sight (LOS)). The $\bar{x}_{\text{HI}}$ is the global neutral fraction and $ \bar{\delta T_b}$ is defined as
\be
\bar{\delta T_b} \simeq (28 \text{mK}) (\frac{\Omega_b h^2}{0.02}) \sqrt{\frac{0.15}{\Omega_m h^2}\frac{1+z}{10}}.
\ee
Eventually, the power spectrum of 21-cm brightness can be written as
\begin{equation}
\begin{aligned}
P_{\delta T_{b}}(k,z) &= \bar{\delta T_b}^2 \bar{x}^2_{\text{HI}}\left[P_{\delta\rho _{\text{HI}}, \delta\rho_{\text{HI}}}(k,z)\right. \\
  &+\left. 2 P_{\delta\rho_{\text{HI}}, \delta\rho_{\text{H}}}\mu^2_k + P_{\delta\rho_{\text{H}}, \delta \rho_{\text{H}}}(k,z) \mu^4 _k\right],
\end{aligned}
\end{equation}
where $P_{x, y}(k,z)$ is the power spectrum of $x$ and $y$ components ($x$ and $y$ each can be neutral and total Hydrogen). Here we assumed that in large scales and higher redshifts the total Hydrogen density traces the dark matter with a very good approximation ($\delta \rho_{\text{H}} \sim \delta \rho_{m}$).
Now, we define the Lagrangian bias corresponding to neutral Hydrogen, ionized Hydrogen and ionized ratio  as $b^{\text{L}}_{a}   =  \delta _{a} (k) / \delta _{\rho_m}$ where $a = \rho_{\text{HI}}, \rho_{\text{HII}}, x_{\text{HII}}$, and $\rho_m$ is the total matter density.
Therefore, equation (\ref{eq:delta_t_1}) and the brightness power spectrum can be written as
\be \label{eq:delta_t_2}
\delta T_b = \bar{\delta T_b}\bar{x}_{\text{HI}}\left(b^{\text{E}}_{\rho _{\text{HI}}} +  \mu^2_k\right)\delta\rho_{m},
\ee
and
\be
P_{\delta T_b} (k,z) = \bar{\delta T_b}^2 \bar{x}^2_{\text{HI}}\left [b^{\text{E}}_{\rho_{\text{HI}}} + \mu_k^2\right]^2P_{mm}(k,z),
\ee
where $b^{\text{E}}_{\rho _{\text{HI}}}\simeq 1+b^{\text{L}}_{\rho _{\text{HI}}}$ is the Eulerian bias, (superscript $E/L$ indicate the Eulerian/Lagrangian bias) and $P_{mm}(k,z)$ is the matter power spectrum in linear regime. Note that the neutral Hydrogen density bias is related to the ionized density as \cite{Mao:2013yaa}
\be\label{eq:b3}
b^{\text{L}}_{\rho_{\text{HI}}}  = \frac{1-\bar{x}_{\text{HII}} b^{\text{L}}_{\rho\text{HII}}}{\bar{x}_{\text{HI}}},
\ee
in which $\bar{x}_{\text{HII}}$ is the ionized fraction.
\cite{Furlanetto:2004nh} shows that the Excursion Set Model for Reionization epoch  (ESMR) works for patchy star forming regions, where $\overline{x}_{\text{HII}} = 1- \overline{x}_{\text{HI}}$. By the definition of $b^{\text{L}}_{x_{\text{HII}}} = \delta_{x_{\text{HII}}} /\delta_m$, we will have
\be\label{eq:b4}
b^{\text{L}}_{ x_{\text{HII}}} \simeq b^{\text{L}}_{\rho\text{HII}}-1.
\ee
Now to find ionized fraction bias we make a basic ansatz that the ionized fraction within spherical volume with radius $R_{0}$ is proportional to the number of ionizing photons produced by sources within that volume, so we simply obtain \cite{DAloisio:2013mgn,DAloisio:2012ioh}
\be
x_{\text{HII}}(M_{\text{min}}, r,z, R_{0})= f_{\text{coll}}(M_{\text{min}}, r,z, R_{0})\zeta,
\ee	
where $\zeta$ is the ionizing efficiency, and $f_{\text{coll}}(M_{\text{min}}, r,z, R_{0})$ is the collapsed mass fraction of volume $R_{0}$ into luminous sources. Here $M_{\text{min}}$ is the mass corresponding to a virial temperature of $10^{4}\text{K}$, so if we assume that the efficiency depends only on the redshift and is independent of position, $\zeta=\zeta(z)$, then the ionized fraction contrast of volume $R_{0}$ will be
\be\label{eq:15}
1+\delta_{x_{\text{HII}}} = \frac{F_{\text{coll}}( > M_{\text{min}}, r, z, R_{0})}{\overline{F}_{\text{coll}}(> M_{\text{min}},z)},
\ee
where $\overline{F}_{\text{coll}}(> M_{\text{min}},z)$ is the mean collapsed fraction of masses larger than $M_{\text{min}}$ at $z$, which is independent of the smoothing scale $R_0$. In the appendix \ref{app} we will discuss the ionized fraction bias in the context of excursion set theory EST \cite{Bond:1990iw,Nikakhtar:2016bju}, where we study the non-Markov effects on the bias parameter \cite{Musso:2012ch}.

In the case of scale-dependent halo bias in form of equation (\ref{eq:kbias}) and by using equations (\ref{eq:b3}, \ref{eq:b4}, \ref{eq:kbias2}) respectively, one simply finds
\be\label{eq:b5}
b^{\text{E}}_{\rho_{\text{HI}}}(k)  =2 -  \frac{\bar{x}_{\text{HII}} }{\bar{x}_{\text{HI}}}b^{\text{L}}_{{x}_{\text{HII}}}- g(k)\frac{\bar{x}_{\text{HII}} }{\bar{x}_{\text{HI}}}b^{\text{L}}_{{x}_{\text{HII}}},
\ee
where $g(k)$ is a scale dependent function which can be raised from PNG or non-Markov effects. In this work, for the matter of simplicity we put scale independent halo bias $b^{\text{L}}_{h}=1$ and $ \frac{\bar{x}_{\text{HII}} }{\bar{x}_{\text{HI}}}b^{\text{L}}_{{x}_{\text{HII}}}=1$. Note that this simplification does not change the main results and proposal of this work.  In the next section we will discuss long mode modulation effect on angular power spectrum.


\section{Angular power spectrum of modulated Hydrogen distribution}
\label{sec:angular}
\begin{figure}
	\includegraphics[width=\columnwidth]{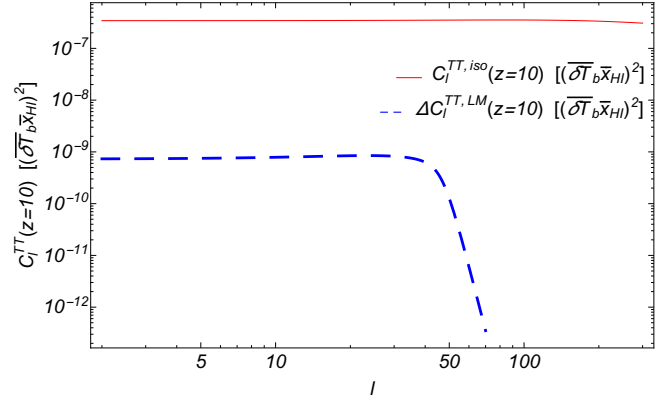}
    \caption{The angular power spectrum ${C}^{TT, \text{iso}}_l$ in units of [${(\bar{\delta T_b}\bar{x}_{\text{HI}})}^{2}$] (red solid line) versus multipole $l$ for 21-cm brightness temperature at $z=10$. The difference of the angular power spectrum due to long mode modulation $\Delta C^{TT, \text{LM}}_l$ at $z=10$ (Blue dashed) is compared.}
    \label{fig:dipole_z10_new}
\end{figure}

\begin{figure}
	\includegraphics[width=\columnwidth]{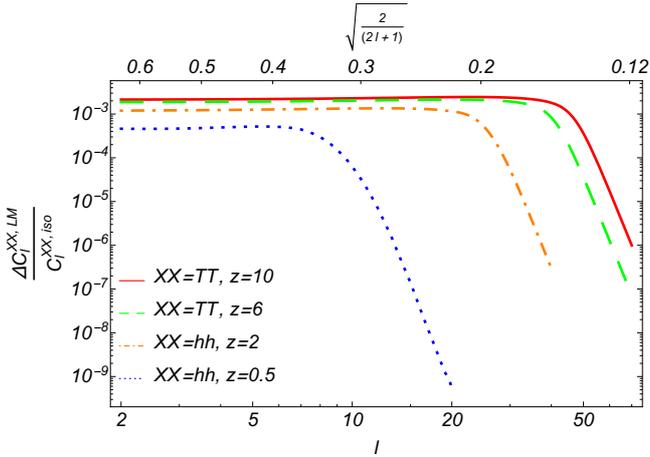}
    \caption{The difference of the angular power spectrum due to long mode modulation for 21-cm brightness temperature with respect to ${C}^{TT, \text{iso}}_l$ ($\Delta C^{TT, \text{LM}}_l/{C}^{TT, \text{iso}}_l$) is plotted versus multipole $l$  at $z=10$ (red solid line) and $z=6$ (green dashed) and for dark matter halo distribution ($\Delta C^{hh, \text{LM}}_l/{C}^{hh, \text{iso}}_l$) at $z=2$ (orange dot dashed) and $z=0.5$ (blue dotted). {The upper $x$-axis shows $\sqrt{{2}/({2l+1})}$ as the ratio of cosmic variance uncertainty on angular power spectrum $\Delta C_l^{\text{cv}}$ to the angular power spectrum $C_{l}$ (see appendix (\ref{app2}) for more information).}  In each multipole the signal $y$-axis must be compared with upper $x$-axis for detection feasibility. }
    \label{fig:dipole_diffrent_redshifts_new}
\end{figure}

In this section we are going to investigate the effect of long mode modulation on the angular power spectrum of matter perturbation. It is known that the Poisson equation in Fourier space relates the matter density to the gravitational potential $k^2\Phi(k)=4\pi G \rho \delta(k)a^2$, and the potential is related to the curvature perturbation via transfer function $T(k)$ and growth function $D(z)$ (i.e. $\Phi(k) = \frac{3}{5} T(k)D(z)(1+z) {\cal{R}}(k)$), therefore the matter density perturbation is related to the curvature as
\be \label{eq:poissonR}
\delta({k})=\frac{2}{5}\frac{T(k)D(z)k^2}{\Omega_m H_0^2}{\cal{R}}({k}).
\ee
{Now the real space 21-cm brightness temperature fluctuations can be expressed in terms of curvature perturbation as}
\begin{equation}
\begin{aligned}
\tilde{\delta T}_b(\hat{n})&=\bar{\delta T_b}\bar{x}_{\text{HI}} \frac{2D(z)}{5\Omega_m H_0^2}\int  \frac{k^2dk d\Omega_k}{(2\pi)^{3/2}}b^{\text{E}}_{\rho _{\text{HI}}} e^{i\vec{k}.\vec{r}}\\
 &\times k^2 T(k) \left [{\cal{R}}^{\text{hi}} + {\cal{R}}^{\text{lo}}+\frac{iA_{\cal{R}}}{r_{\text{LS}}}\frac{\partial}{\partial k_z} {\cal{R}}^{\text{lo}}\right],
\end{aligned}
\end{equation}
where we omit the $k$ in ${\cal{R}}(k)$ for simplicity in notation. {Note that here we neglect redshift space distortion effect. This approximation is based on the  assumption that the RSD angle averaged value in large scales is almost scale independent \cite{Mao:2011xp}. Accordingly, the main result of this work which is a comparison of the modulated angular power spectrum with respect to the standard $\Lambda$CDM case remains unchanged. So for simplicity, and to keep the main point clear, we neglect the redshift space distortion effect. Now we can expand the plane wave in terms of spherical Bessel function} and spherical harmonics, which gives
\begin{equation} \label{eq:integration}
\begin{aligned}
\tilde{\delta T}_b(\hat{n})&=  \bar{\delta T_b}\bar{x}_{\text{HI}} \frac{2}{5}\sqrt{\frac{2}{\pi}} \frac{D(z)}{\Omega_m H_0^2}\int dk k^3 T(k)b^{\text{E}}_{\rho _{\text{HI}}}  \\
&\times \sum_{lm}\left[{\cal{R}}^{\text{hi}}_{lm} + {\cal{R}}^{\text{lo}}_{lm}\left(1+A_{\cal{R}} \frac{\vec{r}}{{r}_{\text{LS}}}.\hat{p} \right)\right] j_{l}(kr)Y_{lm}(\hat{n}),
\end{aligned}
\end{equation}
where
\be
{\cal{R}}^{\text{hi}/\text{lo}}_{lm}(k) = i^l k \int d\Omega_k {\cal{R}}^{\text{hi}/\text{lo}}(k)Y^*_{lm}(\hat{k}).
\ee
For notation simplicity we omit the subscript $b$ hereafter \footnote{We should note that for derivation of equation (\ref{eq:integration}), we use the integration by parts for the term  $\frac{\partial}{\partial k_z} {\cal{R}}^{\text{lo}}$. Now we use the approximation $|i\frac{A_{\cal{R}}}{r_{\text{LS}}} \frac{k_z}{k} \frac{\partial}{\partial k} (k^2T(k))| \ll k^2T(k)$  which is equivalent to $2A_{{\cal{R}}}\chi(z)/( l\times r_{\text{LS}})\ll 1$ where $\chi(z)$ is the comoving distance to the observing redshift $z$ and $l$ is the observing multipole. This approximation allows us to bring back the real space relation $\vec{r}.\hat{p}$ in equation (\ref{eq:integration}) }. Now expanding the 21-cm brightness signal fluctuations in terms of spherical harmonics $\delta \tilde{T}(\hat{n}) = \sum_{lm} \tilde{\delta T}_{lm} Y_{lm}(\hat{n})$, we get
\begin{equation}
\begin{aligned}
\tilde{\delta T}_{lm}&=\bar{\delta T_b}\bar{x}_{\text{HI}}\frac{2}{5}\sqrt{\frac{2}{\pi}} \frac{D(z)}{\Omega_m H_0^2}\left[\int dk k^3 T(k)b^{\text{E}}_{\rho _{\text{HI}}} {({\cal{R}}^{\text{hi}}_{lm}+{\cal{R}}^{\text{lo}}_{lm} )j_l(kr)}\right. \\
&+\left. A_{\cal{R}}\frac{r}{r_{\text{LS}}}\int dk k^3 T(k)\sum_{l'm'}{\cal{R}}^{\text{lo}}_{l'm'} j_{l}(kr)\xi^0_{lml'm'}\right],
\end{aligned}
\end{equation}
where
\be
\xi^M_{lml'm'} \equiv \sqrt{\frac{4\pi}{3}}\int Y^*_{lm}(\hat{n})Y_{l'm'}(\hat{n})Y_{1M}(\hat{n})d\Omega_{\hat{n}}.
\ee
 As we have shown above, the preferred direction is integrated out, consequently the modulation direction cannot be found in our method. In what follows we extensively follow the notation of \cite{Zibin:2015ccn} to find the angular power spectrum of 21-cm brightness signal in the anisotropic case.
 In this direction, it was shown that the multipole covariance in its most general form given a polar (m = 0) modulation can be written in the form of
 ${\langle a_{lm} a^* _{l'm'} \rangle = {C}_l\delta_{ll'}\delta_{mm'}+\frac{1}{2}\delta C_{ll'}\Delta X\xi^0_{lml'm'}}\nonumber$.
Here ${C}_l\delta_{ll'}\delta_{mm'}$ refers to isotropic part of covariance matrix and the non-isotropic part defined as $\delta C_{ll'}$. This formulation works where the anisotropic power spectrum depends on some parameter $X$, linearly \cite{Zibin:2015ccn}. This parameter characterizes the anisotropy by $X(\hat{n})=X_{0}(1+\Delta X/X_{0})cos(\theta)$ and  is related to anisotropic part by $\delta C_{ll'} = dC_l/dX + dC_{l'}/dX$, \cite{Moss:2010qa}.
Accordingly, the covariance matrix of 21-cm brightness temperature fluctuations becomes
\be \label{eq:clmodul}
\langle \tilde{\delta T}_{lm} \tilde{\delta T}^* _{l'm'} \rangle = {C}^{\text{TT},\text{iso}}_l\delta_{ll'}\delta_{mm'}+\frac{1}{2}\delta C^{\text{TT}}_{ll'}\Delta X\xi^0_{lml'm'},
\ee
{ where ${C}^{TT, \text{iso}}_l$ is the isotropic part of angular power spectrum (the angular power spectrum in the absence of long mode modulation)}. The second term in RHS of equation (\ref{eq:clmodul}) is the correction term introduced by the modulated power spectrum {in which $\Delta X = A_{\cal{R}}$}. {Finally, as \cite{Zibin:2015ccn} mentioned the angular power spectrum in the presence of long mode modulation can be written as $C_{l}^{TT}= {C}^{TT, \text{iso}}_l +\Delta C^{TT, \text{LM}}_{l}$ in which  $\Delta C^{TT, \text{LM}}_{l}=\frac{1}{2}A_{\cal{R}}\delta C^{TT}_{ll}$} (note that the label LM on angular power spectrum, indicates long mode modulation), so
\begin{equation} \label{eq:isotropic}
\begin{aligned}
{C}^{TT, \text{iso}}_l&={(\bar{\delta T_b}\bar{x}_{\text{HI}})}^{2}\frac{16\pi}{25}(\frac{D(z)}{\Omega_m H_0^2})^2\\&\times\int \frac{dk}{k} k^4 T^2(k){\cal{P}}_{\cal{R}}(k){(b^{\text{E}}_{\rho _{\text{HI}}})}^{2}j^2_l(kr),
\end{aligned}
\end{equation}
where ${\cal{P}}_{\cal{R}}$ is the dimensionless power spectrum of curvature perturbation, and
\begin{equation}\label{eq:deltalm}
\begin{aligned}
\Delta C^{TT, \text{LM}}_l&={(\bar{\delta T_b}\bar{x}_{\text{HI}})}^{2}\frac{32\pi}{25}(\frac{D(z)}{\Omega_m H_0^2})^2 A_{\cal{R}}\frac{r}{r_{\text{LS}}} \\&\times\int \frac{dk}{k} k^4 T^2(k){\cal{P}}^{\text{lo}}_{\cal{R}}(k){(b^{\text{E}}_{\rho _{\text{HI}}})}^{2}j^2_l(kr).
\end{aligned}
\end{equation}
Here we assume that neutral density bias has no k-dependence ($b^{\text{E}}_{\rho_{\text{HI}}}  =2 -  \frac{\bar{x}_{\text{HII}} }{\bar{x}_{\text{HI}}}b^{\text{L}}_{{x}_{\text{HII}}}$). As far as the bias parameter is scale independent, the main contribution to the angular power spectrum comes from the matter power spectrum corrections.
~In Fig.\ref{fig:dipole_z10_new}, we plot the angular power spectrum of 21-cm brightness fluctuations versus angular multipole $l$ and also we compare the signal with long mode modulation effect (equation (\ref{eq:deltalm})) in $z=10$. {In order to explicitly show the advantage of using the 21-cm brightness power spectrum in comparison to the late time dark matter tracer's power (e.g. galaxies), we study the halo-halo power spectrum in this respect.  In Fig.\ref{fig:dipole_diffrent_redshifts_new}, we compare the fractional difference of angular power spectrum of 21-cm brightness and dark matter halo distribution in different redshifts (for details see caption of the figures). The dark matter halo distribution angular power spectrum ${C}^{hh}_l$ is obtained
\begin{equation} \label{eq:halo-power}
{C}^{hh}_l=\frac{16\pi}{25}(\frac{D(z)}{\Omega_m H_0^2})^2\int \frac{dk}{k} k^4 T^2(k){\cal{P}}_{\cal{R}}(k){b^{\text{E}}_{h}}^{2}j^2_l(kr),
\end{equation}
where $b^{\text{E}}_{h}$ is the halo bias term. This relation is obtained by the fact that the halo density contrast $\delta_h$ is related to the matter density contrast $\delta_m$ approximately via the Eulerian bias $\delta_h\simeq b^{\text{E}}_{h} \delta_m$. On the other hand the matter density contrast is related to curvature perturbations by equation (\ref{eq:poissonR})}. \footnote{ All the corrections to the halo-halo matter power spectrum due to long mode modulation and non-Gaussianity are calculated  on the basis of equation (\ref{eq:halo-power}), similarly.} In Fig. \ref{fig:dipole_diffrent_redshifts_new}, we show that the halo-halo power spectrum changes due to long mode, dies off in smaller multipoles in comparison to the 21-cm brightness power spectrum. This is reasonable as for a fixed wavenumber, the power is changed in smaller multipoles, when we probe lower redshifts.  As one can see in Fig. \ref{fig:dipole_z10_new}, \ref{fig:dipole_diffrent_redshifts_new}, the long mode modulation dies at large multipoles similar to CMB observations, but as \cite{{Zibin:2015ccn}} has made the cut off by a fixed $k$ ($k_c$) instead of $l$, this suppression cut off will shift to lower multipoles by decreasing redshift. In other words, the cosmic variance problem is more dominant in lower redshifts. {As we discuss in appendix (\ref{app2}), at low multipoles the main source of uncertainty will be the cosmic variance $\Delta C_l^{\text{cv}}=\sqrt{{2}/({2l+1})}C_{l}$.
In order to have a general comparison between the change in angular power spectrum due to long mode modulation and the cosmic variance, the upper $x$-axis in Fig. \ref{fig:dipole_diffrent_redshifts_new} is labeled  $\sqrt{\frac{2}{2l+1}}$. This is the ratio of cosmic variance to the $C_{l}$ (In each multipole the signal $y$-axis must be compared with upper $x$-axis for detection feasibility).} Furthermore, due to the suppression term $r/r_{\text{LS}}$ in equation (\ref{eq:curvature_lm}), the effective amplitude of long mode modulation will reduce in lower redshifts. Although the observing of {dipole modulation} in 21-cm temperature brightness map of EoR is more difficult than CMB map, but eventually it's more promising and reachable than low redshift galaxy surveys. {Another important point that we should {emphasise} here is the issue of the cosmic variance. As one can see in Fig. \ref{fig:dipole_diffrent_redshifts_new} the long mode modulation feature on angular power spectrum is in cosmic variance limit for all scales (for more information see appendix (\ref{app2})) and therefore cannot be observed. However, as we will show in next section, local type PNG can enhance this feature in the angular power spectrum of 21-cm signal at high redshifts to the values which are not cosmic variance limited.}

In the next section we will study the effect of local type PNG on bias parameter and angular power spectrum.

\section{Non-Gaussianity: the bias and angular power spectrum}
\label{sec:PNGbias}
In this section we investigate the effect of power asymmetry in the presence of primordial local non-Gaussianity.
As indicated in the introduction, the long mode modulation can be accompanied by local non-Gaussianity which is related to the long-to-short mode coupling. In this direction, we study the local $f_{\text{NL}}$ type non-Gaussianity.
The primordial Bardeen potential can be expressed in terms of local non-Gaussian corrections as
\be
\Phi=\phi_g+f_{\text{NL}}\phi_g^2,
\ee
where the non Gaussian pre-factors $f_{\text{NL}}$ is a constant and $\phi_g$ is a Gaussian field. We assume that $\langle\phi_g\rangle=0$. In order to calculate the bias we use the peak-background splitting method \cite{Sheth:1999mn}. The Bardeen potential can be written in terms of short mode $\phi_s$ and long mode $\phi_l$ Gaussian fields ($\phi_g=\phi_s+\phi_l$),
\begin{equation}\label{Eq:bardeen}
\Phi=\phi_s+\phi_l+f_{\text{NL}}\phi_s^2+2f_{\text{NL}}\phi_s\phi_l+f_{\text{NL}}\phi_l^2. 
\end{equation}
Assuming that $\phi_s\gg\phi_l$, we can rewrite the equation (\ref{Eq:bardeen}) up to the first order of $\phi_l$
\be
\Phi=\phi_l+X_1\phi_s+X_2\phi_s^2,
\ee
where $X_1=1+2f_{\text{NL}}\phi_l$ and $X_2=f_{\text{NL}}$.
Now we want to find the effect of non-Gaussian terms on halo bias defined $b(M,k,z)=\delta_h(M,k,z)/\delta_m(z)$
where $\delta_h$ is the halo number density contrast and $\delta_m$ is the linear matter density
which is identical to the long mode matter density contrast $\delta_l$. We should note that  bias in its general form, depends on the mass of dark matter halo, the scale of observation, and redshift. Density contrast in the late time can be related to the $\phi_l$ in the early universe by Poisson equation $
\delta_l={\cal{M}}(k,z)\phi_l$, where ${\cal{M}}(k,z)=3k^2 T(k) D(z) /5H_0^2\Omega_m$. 
 Note that the bias is a scale dependent quantity due to the non-Gaussian effects, which is discussed in \cite{Dalal:2007cu}. The halo density contrast with PNG corrections is defined as
\be \label{Eq:halodensity}
\delta_h=\frac{n(M,z,X_1,X_2;\delta_l)-\bar{n}(M,z)}{\bar{n}(M,z)},
\ee
where $\bar{n}(M,z)$ is the number density of structures in a mass range of $M$ and $M+dM$ in redshift $z$. By Taylor expanding the modified number density per mass $n(M,z,X_1,X_2;\delta_l)$ in equation (\ref{Eq:halodensity}) in the presence of long mode $\delta_l$ and non-Gaussian terms, we can find the bias parameter due to the definition of halo bias (see appendix (\ref{app})) as
\be
\bar{b}^{\text{L}}_h=\bar{b}^{(g,\text{L})}_h+\frac{\beta_2 f_{\text{NL}}}{{\cal{M}}(k,z)},
\ee
where $\bar{b}^{(g,\text{L})}_h\equiv \partial (\ln\bar{n})/\partial \delta_l$ is the Gaussian Lagrangian bias which depends on the universality function. $\beta_2$ is defined as derivative of the number density with respect to $X_1$
\be
\beta_2=2\frac{\partial\ln \bar{n}}{\partial X_1},
\ee
and therefore
\begin{equation}
\bar{b}^{\text{L}}_h (f_{\text{NL}}) =  \bar{b}^{(\text{L},g)}_h + 2f_{\text{NL}} {\cal{M}}^{-1}(k,z) \bar{b}^{(\text{L},g)}_h\delta_c(z),
\end{equation}
where $\delta_c=1.686$ is the critical density. So according to equation (\ref{eq:b5}), we can extend the definition of halo bias to neutral Hydrogen bias. In this case the Eulerian neutral Hydrogen bias $b^{E}_{\rho_{HI}}$ can be expressed as
\be \label{eq:Hydrogenbias}
b^{\text{E}}_{\rho_{\text{HI}}}(k)  =2 -  \frac{\bar{x}_{\text{HII}} }{\bar{x}_{\text{HI}}}b^{\text{L}}_{{x}_{\text{HII}}}- 2f_{\text{NL}} {\cal{M}}^{-1}(k,z)\delta_c(z)\frac{\bar{x}_{\text{HII}} }{\bar{x}_{\text{HI}}}b^{\text{L}}_{{x}_{\text{HII}}}.
\ee
\begin{figure}
	\includegraphics[width=\columnwidth]{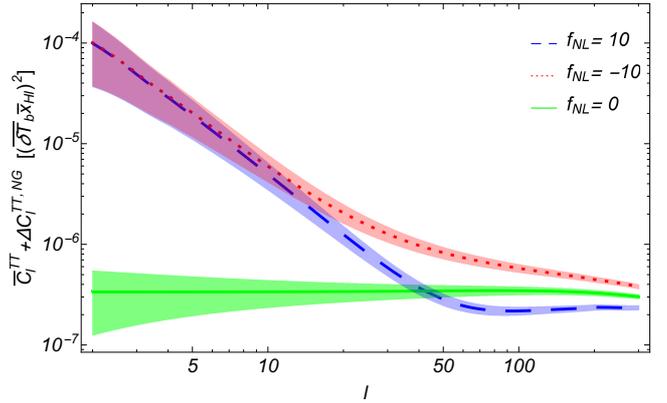}
    \caption{The angular power spectrum of  21-cm brightness temperature including non-Gaussianity $\bar{C}^{TT}_l + \Delta C^{TT, \text{NG}}_l$  versus multipole $l$ at $z=10$ for $f_{\text{NL}}=0$ (green solid line), $f_{\text{NL}}=+10$ (blue dashed) and $f_{\text{NL}}=-10$ (red dotted). {The shaded regions correspond to the cosmic variance uncertainty area ($C_{l}\pm\Delta C_l^{\text{cv}}=(1\pm \sqrt{{2}/({2l+1})})C_{l}$). See appendix (\ref{app2}) for more information.}}
    \label{fig:NG_z10_new}
\end{figure}
\begin{figure}
	\includegraphics[width=\columnwidth]{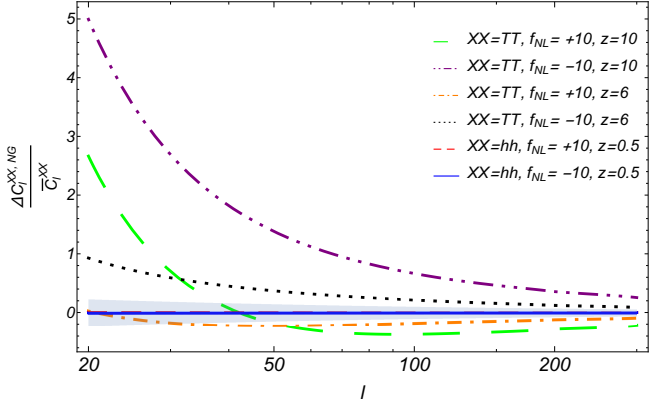}
    \caption{The difference of the angular power spectrum due to non-Gaussianity with respect to $\bar{C}_l$ for 21-cm brightness temperature ($\Delta C^{TT, \text{NG}}_l/\bar{C}^{TT}_l$) and for dark matter halo distribution ($\Delta C^{hh, \text{NG}}_l/\bar{C}^{hh}_l$) is plotted versus multipole $l$ at different redshifts with $f_{\text{NL}}=\pm10$.
     Long-dashed (green) and dot-dot-dashed (purple) lines correspond to  21-cm brightness temperature with  $f_{\text{NL}}=+10$ and $f_{\text{NL}}=-10$ at $z=10$, dot-dashed (orange) and dotted (black)  lines are for  21-cm brightness temperature with  $f_{\text{NL}}=+10$ and $f_{\text{NL}}=-10$ at $z=6$ and finally short-dashed (red) and solid (blue) lines  are for dark matter halo distribution with  $f_{\text{NL}}=+10$ and $f_{\text{NL}}=-10$ at $z=0.5$ respectively. {The shaded region corresponds to the cosmic variance uncertainty area ($\pm\Delta C_l^{\text{cv}}/C_{l}=\pm \sqrt{\frac{2}{2l+1}}$). See appendix (\ref{app2}) for more information.}}
    \label{fig:NG_diffrent_redshifts_1_new}
\end{figure}
\begin{figure}
	\includegraphics[width=\columnwidth]{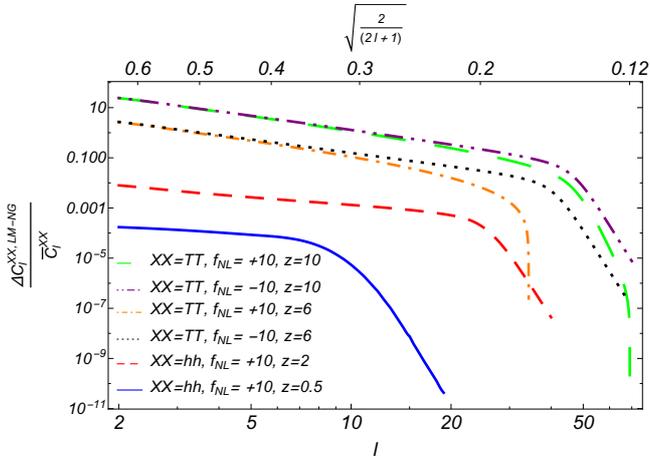}
	\caption{The difference of the angular power spectrum due to LM-NG term $\Delta C^{\text{LM-NG}}_l$ introduced in equation (\ref{eq:cl-ng-lm}) with respect to $\bar{C}_l$ for 21-cm brightness temperature $\Delta C^{TT, \text{LM-NG}}_l/\bar{C}^{TT}_l$  and for dark matter halo distribution  is plotted versus multipole $l$  at different redshifts with $f_{\text{NL}}=\pm10$. Long-dashed (green) and dot-dot-dashed (purple) lines are for 21-cm brightness temperature with $f_{\text{NL}}=+10$ and $f_{\text{NL}}=-10$ at $z=10$, dot-dashed (orange) and dotted (black)  lines are for  21-cm brightness temperature with  $f_{\text{NL}}=+10$ and $f_{\text{NL}}=-10$ at $z=6$ and finally short-dashed (red) and solid (blue) lines  are for dark matter halo distribution with  $f_{\text{NL}}=+10$ at $z=2$ and $z=0.5$ respectively. {The upper $x$-axis shows $\sqrt{{2}/({2l+1})}$ as the ratio of cosmic variance uncertainty on angular power spectrum $\Delta C_l^{\text{cv}}$ to the angular power spectrum $C_{l}$ (see appendix (\ref{app2}) for more information). In each multipole the signal $y$-axis must be compared with upper $x$-axis for detection feasibility.}}
	\label{fig:ng_dipole_diffrent_redshifts_new}
\end{figure}
\begin{figure}
	\includegraphics[width=\columnwidth]{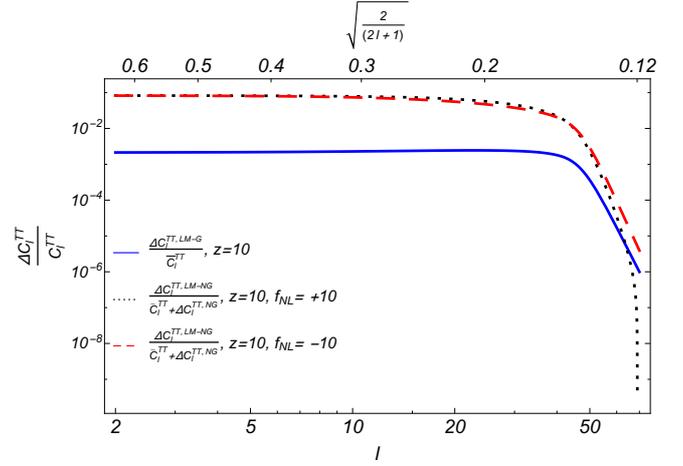}
	\caption{The ratio of the angular power spectrum due to long mode modulation $\Delta C^{TT, \text{LM-G}}_l$ with respect to $\bar{C}^{TT}_l$ (in absence of non-Gaussianity) versus multipole $l$  for 21-cm brightness temperature $\Delta C^{TT, \text{LM-G}}_l/\bar{C}^{TT}_l$ at $z=10$ (blue solid line). Also, the difference of the angular power spectrum due to LM-NG term (equation (\ref{eq:cl-ng-lm})) $\Delta C^{TT, \text{LM-NG}}_l$ with respect to $\bar{C}^{TT}_l+ \Delta C^{TT, \text{NG}}_l$ with $f_{\text{NL}}=+10$ (dotted black) and $f_{NL}=-10$ (dashed red) is compared. { For the cosmic variance see the captions of Fig. \ref{fig:dipole_diffrent_redshifts_new} and Fig. \ref{fig:ng_dipole_diffrent_redshifts_new}.}}
	\label{fig:z10_ng_effect_on_lm_plot}
\end{figure}
\begin{figure}
	\includegraphics[width=\columnwidth]{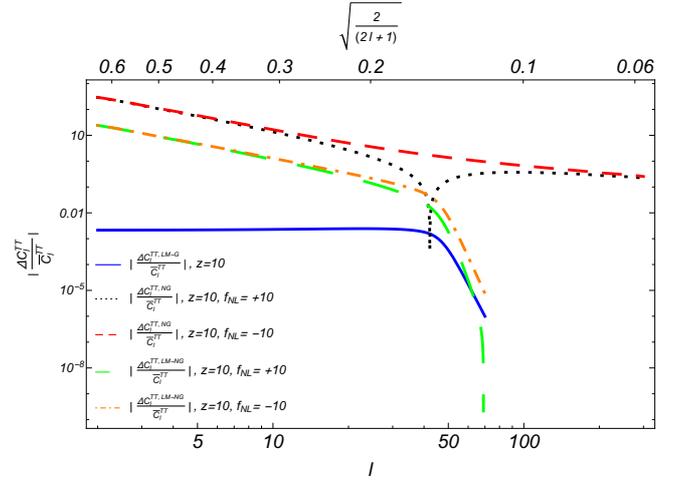}
    \caption{Absolute value of the difference of the angular power spectrum due to each term in equations (\ref{eq:34}-\ref{eq:cl-ng-lm}) with respect to $\bar{C}^{TT}_l$ versus multipole $l$  for 21-cm brightness temperature $\Delta C^{TT}_l/\bar{C}^{TT}_l$ at $z=10$. Dotted (black) and short-dashed (red) lines are for $\Delta C^{TT, \text{NG}}_l/\bar{C}^{TT}_l$ with $f_{\text{NL}}=+10$ and $f_{\text{NL}}=-10$, long-dashed (green) and dot-dashed (orange) lines are for $\Delta C^{TT, \text{LM-NG}}_l/\bar{C}^{TT}_l$ with $f_{\text{NL}}=+10$ and $f_{NL}=-10$ and finally blue solid line is for $\Delta C^{TT, \text{LM-G}}_l/\bar{C}^{TT}_l$ term respectively. {For the cosmic variance see the captions of Fig. \ref{fig:dipole_diffrent_redshifts_new} and Fig. \ref{fig:ng_dipole_diffrent_redshifts_new}.}}
    \label{fig:z10_final_plot}
\end{figure}
Now it is straightforward to incorporate the effect of PNG via the bias parameter in the definition of the angular power spectrum of 21-cm brightness temperature fluctuations. One of the important studies in this framework was done by \cite{Slosar:2008hx}, which investigate the effect of PNG on lower redshifts to compare the results of PNG corrections with the SDSS data.  Our PNG correction results  is compatible with those of \cite{Slosar:2008hx} for low redshifts and \cite{Cooray:2008eb} for EoR (In Fig. \ref{fig:NG_z10_new} and Fig. \ref{fig:NG_diffrent_redshifts_1_new}, we plot the effect of PNG on angular power spectrums both for low and high redshifts, where we neglect the redshift space distortion and non-linear effects). However, in this work we study the angular power spectrum of temperature brightness fluctuations in presence of both PNG and long mode modulation as well. In the presence of long mode modulation, {as we showed, the angular power spectrum contains an isotropic term and a modified term due to LM modulation. However in the presence of local type non-Gaussianity by using equations (\ref{eq:isotropic}, \ref{eq:deltalm}, \ref{eq:Hydrogenbias}), both isotropic and long mode modulated angular power spectrum can be divided to Gaussian and non-Gaussian terms as ${C}^{TT, \text{iso}}_l= \bar{C}^{TT}_l + \Delta C^{TT, \text{NG}}_l $ and $\Delta C^{TT, \text{LM}}_{l}=\Delta C^{TT, \text{LM-G}}_l+\Delta C^{TT, \text{LM-NG}}_l$. Therefore the angular power spectrum becomes}
\begin{equation}\label{eq:34}
C^{TT}_l= \bar{C}^{TT}_l + \Delta C^{TT, \text{LM-G}}_l + \Delta C^{TT, \text{NG}}_l +  \Delta C^{TT, \text{LM-NG}}_l,
\end{equation}
 where $\bar{C}^{TT}_l$ is the angular power spectrum in absence of both long mode modulation and local type non-Gaussianity obtained as
\begin{equation}
\begin{aligned}
\bar{C}^{TT}_l=& {\left(\bar{\delta T_b}\bar{x}_{\text{HI}}\left(2 -  \frac{\bar{x}_{\text{HII}} }{\bar{x}_{\text{HI}}}b^{\text{L}}_{{x}_{\text{HII}}}\right)\right)}^{2}\frac{16\pi}{25}\left(\frac{D(z)}{\Omega_m H_0^2}\right)^2\\
&\times\int \frac{dk}{k} k^4 T^2(k){\cal{P}}_{\cal{R}}(k)j^2_l(kr),
\end{aligned}
 \end{equation}
and $\Delta C^{TT, \text{LM-G}}_l$ is the term which is related to the pure long mode modulation effect
\begin{equation}
\begin{aligned}
\Delta C^{TT, \text{LM-G}}_l=& {\left(\bar{\delta T_b}\bar{x}_{\text{HI}}\left(2 -  \frac{\bar{x}_{\text{HII}} }{\bar{x}_{\text{HI}}}b^{\text{L}}_{{x}_{\text{HII}}}\right)\right)}^{2}\frac{32\pi}{25}\left(\frac{D(z)}{\Omega_m H_0^2}\right)^2 \\
&\times A_{\cal{R}}\frac{r}{r_{\text{LS}}}\int \frac{dk}{k} k^4 T^2(k){\cal{P}}^{\text{lo}}_{\cal{R}}(k)j^2_l(kr),
\end{aligned}
\end{equation}
and  $\Delta C^{TT, \text{NG}}_l$ is the effect of PNG scale-dependent bias term on standard angular power spectrum
\begin{equation}
\begin{aligned} \label{eq:cl-ng}
\Delta C^{TT, \text{NG}}_l&= {\left(\bar{\delta T_b}\bar{x}_{\text{HI}}\right)}^{2}\frac{16\pi}{25}\left(\frac{D(z)}{\Omega_m H_0^2}\right)^2\\
&\times\int \frac{dk}{k} k^4 T^2(k){\cal{P}}_{\cal{R}}(k)\\
&\times\left[-4f_{\text{NL}}\delta_c {\cal{M}}^{-1}(k,z)\frac{\bar{x}_{\text{HII}} }{\bar{x}_{\text{HI}}}b^{\text{L}}_{{x}_{\text{HII}}}\left(2 -  \frac{\bar{x}_{\text{HII}} }{\bar{x}_{\text{HI}}}b^{\text{L}}_{{x}_{\text{HII}}}\right)\right.\\
&+\left.{\left(2f_{\text{NL}}\delta_c {\cal{M}}^{-1}(k,z)\frac{\bar{x}_{\text{HII}} }{\bar{x}_{\text{HI}}}b^{\text{L}}_{{x}_{\text{HII}}}\right)}^{2}\right]j^2_l(kr),
\end{aligned}
\end{equation}
and $\Delta C^{TT, \text{LM-NG}}_l$ is the effect of PNG scale-dependent bias on long mode modulated term
\begin{equation}
\begin{aligned} \label{eq:cl-ng-lm}
\Delta C^{TT, \text{LM-NG}}_l&= {(\bar{\delta T_b}\bar{x}_{\text{HI}})}^{2}\frac{32\pi}{25}\left(\frac{D(z)}{\Omega_m H_0^2}\right)^2 \\
&\times A_{\cal{R}}\frac{r}{r_{\text{LS}}}\int \frac{dk}{k} k^4 T^2(k){\cal{P}}^{\text{lo}}_{\cal{R}}(k)\\
&\times\left[-4f_{\text{NL}}\delta_c {\cal{M}}^{-1}(k,z)\frac{\bar{x}_{\text{HII}} }{\bar{x}_{\text{HI}}}b^{\text{L}}_{{x}_{\text{HII}}}\left(2 -  \frac{\bar{x}_{\text{HII}} }{\bar{x}_{\text{HI}}}b^{\text{L}}_{{x}_{\text{HII}}}\right)\right.\\
&+\left. {\left(2f_{\text{NL}} \delta_c {\cal{M}}^{-1}(k,z)\frac{\bar{x}_{\text{HII}} }{\bar{x}_{\text{HI}}}b^{\text{L}}_{{x}_{\text{HII}}}\right)}^{2}\right]j^2_l(kr).
\end{aligned}
\end{equation}
In Fig. \ref{fig:NG_z10_new}, the angular power spectrum of  21-cm brightness temperature including the non-Gaussianity $\bar{C}^{TT}_l + \Delta C^{TT, \text{NG}}_l$ term at $z=10$ for $f_{\text{NL}}=0, \pm10$ is plotted. It is interesting to note that  due to the term ${\cal{M}}^{-1}$ in equation (\ref{eq:cl-ng}) (which is proportional to $k^{-2}T^{-1}(k)$) non-Gaussianity term will have a very large effect on angular power spectra at low multipoles ($ \Delta C^{TT, \text{NG}}_l>\bar{C}^{TT}_l$ at $l\lesssim 50$ with $f_{\text{NL}}=+10$). This is a very interesting {feature} with regards to PNG studies, and it can be used to detect PNG \cite{Lidz:2013tra}.  In Fig. \ref{fig:NG_diffrent_redshifts_1_new}, we plot the ratio of the standard angular power spectrum with PNG correction with respect to the standard angular power spectrum. For redshifts $z=10$ and $z=6$ we plot the 21-cm brightness temperature fluctuations angular power, where for $z=0.5$ the angular power spectrum is for dark matter halo distribution. This plot shows how it is promising to use 21-cm signal power spectrum to find out the PNG effects. In Fig. \ref{fig:ng_dipole_diffrent_redshifts_new} we show the ratio of the PNG-long mode modulated term in angular power spectrum introduced in equation (\ref{eq:cl-ng-lm}) with respect to the standard power. This figure shows that PNG-long mode modulated term with respect to the standard power spectrum, also decreases in lower redshifts. In order to emphasis this result in Fig. \ref{fig:z10_ng_effect_on_lm_plot}, which is an important plot for our proposal, we show that PNG not only enhances the long mode modulation effect (up to 4 order of magnitude), but also enhances its contribution with respect to main terms in angular power spectrum (up to 2 order of magnitude at $z=10$) to an observable signal. Also note that this signal is not cosmic variance limited.
In Fig. \ref{fig:z10_final_plot}, we plot the contribution of each term introduced in this section. {The upper $x$-axis in Figs. \ref{fig:ng_dipole_diffrent_redshifts_new}, \ref{fig:z10_ng_effect_on_lm_plot}, \ref{fig:z10_final_plot} shows $\sqrt{{2}/({2l+1})}$ as the ratio of cosmic variance to the $C_{l}$. In each multipole the signal $y$-axis must be compared with upper $x$-axis for detection feasibility (see appendix (\ref{app2}) for more information).} Summing up the result, we show that if the long mode modulation has a physical origin, it can be enhanced by PNG and it is possible to observe it in future surveys.


\section{Conclusions and Further remarks}
\label{sec:conclusion}
Deviation from the standard picture of inflationary model with nearly Gaussian, nearly scale invariant, isotropic and adiabatic initial conditions can open up a new horizon to the physics of early universe. In this work we study the effect of long mode modulation as a {possible explanation} for CMB {hemispherical power asymmetry} with local type of non-Gaussianity, on the distribution of neutral Hydrogen in EoR. This study is a continuation of the idea of using late time large scale structure observations as a complementary probe for EU physics (In this direction see \cite{Hirata:2009ar,Baghram:2013lxa,Zibin:2014rfa, Hassani:2015zat,Fard:2017oex, Zhai:2017ibd}). {In the present work  it is shown that
although the observation of {dipole modulation} in 21-cm temperature brightness map of EoR is more difficult than CMB map, but eventually it's more promising and more reachable than low redshift galaxy surveys}. It is shown in Fig. \ref{fig:dipole_diffrent_redshifts_new} how two different type of observation predict the large angular scale corrections. Here, we also study the effect of PNG in the presence of long mode modulation. 
 The local type PNG has a $k^{-2}$ dependency on  wavenumber in the bias parameter, therefore the PNG in its local type can enhance the long mode modulation effect on angular power spectrum. As it is discussed, the long mode modulations are companied with local type PNG. Therefore, this seems a useful coincidence which can help to detect this effect.
For further study, it is necessary to introduce a more sophisticated bias model for neutral Hydrogen by using the simulations to pin down the physics of ionized patches and the efficiency function of ionization. On the other hand, more realistic models of non-Markov bias indicate a $k$-dependency which must be taken into account. These corrections may have  drastic effects on the angular power spectrum of 21-cm signal. Another direction for the extension of our analysis is the study of the effect of the RSD corrections on the angular power spectrum. As we discussed in the main text, we neglect the RSD effect due to its nearly constant contribution to the angular power spectrum in large scale. However this is an approximation and further study is needed to incorporate the RSD effect properly.   Summing up this work we should note that  the future observations such as the Square Kilometer Array (SKA) with enough large angle coverage may become plausible candidates to detect the large scale anomalies of initial condition of early universe \cite{Bacon:2018dui}. In the appendix (\ref{app2}) we discuss the characteristics of the SKA project, mainly focusing on the errors introduced on the angular power spectrum. Also we show that with the simple and approximate analysis, the effect of long mode modulation and PNG could be larger than the systematics of SKA. The more detailed analysis of the detectability  of these effects will be the subject of a future work. \\

{Note added: During the final stage of this work the Planck 2018 results. VII. Isotropy and Statistics of the CMB results on  the isotropy of cosmic microwave background radiation and the studies of the CMB anomalies appeared \cite{Akrami:2019bkn}. The main findings of the paper regarding this work is that the hemispherical power asymmetry is still observed in temperature maps. However the E-mode polarization map of CMB does not provide a new evidence for this anomaly. However, the paper concluded that the independent search for CMB anomalies including the hemispherical power asymmetry is encouraged to found out the origin of this features, which could be beyond standard model.}
\section*{Acknowledgements}
We would like to thank Ali Akbar Abolhasani, Mohammad Reza Rahimitabar, Toyokazu Sekiguchi and Ravi K. Sheth for useful comments and fruitful
discussions. We should also thank the anonymous referee for his/her valuable and insightful comments, which help us to improve the manuscript. SK and SB thank the school of Astronomy of Institute for Research in Fundamental Sciences (IPM) for hospitality during this work. SB is partially supported by Abdus Salam International Center of Theoretical Physics (ICTP) under the junior associateship  scheme during this work. This research is supported by Kharazmi University Office of Vice President for Research under Grant No. 41/2042 and Sharif University of Technology Office of Vice President for Research under Grant No. G960202.



%
\appendix

\section{Modulated bias in Excursion set theory}\label{app}

\label{sec:modulatedbias}
 In the first glance, bias is considered as a nuisance parameter, but it can also be studied as a parameter for detecting the anomalous initial condition in the process of collapse. In this appendix we study the ionized fraction bias in the context of Excursion Set Theory (EST).

In the context of the excursion set theory and peak background splitting \cite{Sheth:1999mn}, bias can be defined by the first up-crossing distribution as
\be
b_n = \frac{(-1)^n}{f(s,\delta_c)}\frac{\partial^n f(s,\delta_c)}{\partial \delta_c^n},
\ee
where $f(s,\delta_c)$ is the first up-crossing function, which depends on the variance $s$ and critical density of spherical collapse $\delta_c$. Note that  $n=1$ corresponds to the linear regime Lagrangian bias and for the Markov first up-crossing $f(s,\delta_c)= 1/\sqrt{2\pi} ( \delta_c/s^{3/2}) e^{-\delta^2_c / 2s}$, we will have $b_1= (\nu_c^2 - 1)/\delta_c$, where $\nu = \delta / \sqrt{s}$ is the height parameter.

Now according to equation (\ref{eq:15}), in the excursion set theory we can find mean value of the collapsed fraction of the ionization ratio as
\be
\langle 1+ \delta_{x_{\text{HI}}} \rangle  = \frac{F(s(M_{\text{min}})\mid s_{0},\delta_{0})}{F(s(M_{\text{min}}))},
\ee
in which $s(M_{\text{min}})$ is the variance of density fluctuations related to the minimum mass that can ionize the surroundings with scales larger than the scale $s_0$ under consideration. The function $F$ is the integrated fraction of up-crossing.
Now if we assume that the walks are uncorrelated steps and they are generated by Gaussian initial condition the collapse fraction will be \cite{Mao:2013yaa}
\be
F(s(M_{\text{min}})\mid s_{0},\delta_{0})= \text{erfc}[\frac{\delta_{c}(z)-\delta_{0}}{\sqrt{2(s(M_{\text{min}})-s_{0})}}].
\ee
In order to find the bias parameter in real space, we can simply set $s_{0}=0$ and expand $\langle 1+ \delta_{x_{\text{HI}}} \rangle$ around $\delta_{0}=0$ to obtain
\be\label{eq14}
\frac{F(s(M_{\text{min}})\mid s_{0}=0,\delta_{0})}{F(s(M_{\text{min}}))}= 1 + \sum \frac{{\delta_{0}}^{n}}{n!}b_{ x_{\text{HII}},n},
\ee
and thus the first order bias parameter can be found as
\be
b_{ x_{\text{HII}},1}=\sqrt{\frac{2}{\pi s(M_{\text{min}})}} \exp{(-\frac{{\delta_{c}}^{2}}{2s(M_{\text{min}})})}({\text{erfc}[\frac{\delta_{c}}{\sqrt{2s(M_{\text{min}})}}]})^{-1},
\ee
where $s(M_{min})$ and $\delta_{c}$ vary with redshift. It is seen that similar to the real space bias there is no scale dependency in Fourier space \cite{Musso:2012ch}
\be
b_{ x_{\text{HII}}}(k)\sim b_{ x_{\text{HII}},1}(k)=b_{ x_{\text{HII}},1}.
\ee
Note that these coefficients are pure numbers, independent of wavenumber k, and by definition, independent of $s_0$.

The bias parameter introduced above corresponds to uncorrelated steps in EST approach which refers to a k-space sharp smoothing window  function for density field. For more realistic physical cases, other window functions such as  real space top hat and Gaussian filters are suggested. These filters  lead to the correlated steps in EST approach \cite{Nikakhtar:2018qqg}. Although there is no analytical exact solution for first up-crossing problem with correlated steps, there are some approximate solutions in this case.
Here we find the ionized fraction bias in the context of the correlated steps in EST. Note that in the ionized fraction bias we used  $F(s(M_{\text{min}})\mid s_{0}=0,\delta_{0})$ from EST which is mass fraction of volume $s_{0}$ that collapsed in halos with scales between $s_{0}$ and $s(M_{min})$. \cite{Musso:2012ch} found approximate solutions for correlated steps of halo bias
\be
\langle 1+ \delta_{h} \rangle  = \frac{f(s\mid s_{0},\delta_{0})}{f(s)}= 1 + \sum \frac{{\delta_{0}}^{n}}{n!}b_{h,n},
\ee
in which $f(s\mid s_{0},\delta_{0})$ is the mass fraction of volume $s_{0}$ that collapsed in halos with scales between $s$ and $s+ds$. With this definition in mind, we simply have
\begin{equation}\label{eq18}
\begin{aligned}
F(s_{\text{min}}\mid s_{0},\delta_{0}) =& \int_{s_{0}}^{s_{\text{min}}}f(s\mid s_{0},\delta_{0})ds\\&
=\int_{S_{0}}^{s_{\text{min}}}\langle 1+ \delta_{h} \rangle f(s)ds\\&
=F(s_{\text{min}}) + \sum \frac{{\delta_{0}}^{n}}{n!}\int_{s_{0}}^{s_{\text{min}}}f(s)b_{h,n}(s,s_{0})ds.
\end{aligned}
\end{equation}
Note that we had made no assumption on whether bias parameter has scale dependency or not. By comparing equations (\ref{eq18}) and (\ref{eq14}), one can simply find
\be
b_{ x_{\text{HII}},n}=\frac{\int_{S_{0}}^{s_{\text{min}}}f(s)b_{h,n}(s,S_{0})ds}{F(s_{\text{min}})},
\ee
where
\be\label{eq20}
F(s_{\text{min}})=\int_{0}^{s_{\text{min}}}f(s)ds.
\ee
The same relation can be found for Fourier space bias
\be\label{eq:int}
b_{ x_{\text{HII}},n}(k)=\frac{\int_{0}^{s_{\text{min}}}f(s)b_{h,n}(k)ds}{F(s_{\text{min}})}.
\ee

Now we can use this relations to find ionized fraction bias for top hat and Gaussian filters. As mentioned above, there is no exact solution for these window functions so we use the up-crossing approximation introduced in \cite{Musso:2012ch}. Here we only discuss scale independent part of bias ($b_{10}$)  as below
\be
sf_{\text{up}}(s)=\frac{\nu \exp{(\frac{-\nu^2}{2})} }{2\sqrt{2\pi}}(\frac{1+\text{erf}(\Gamma\nu/\sqrt{2})}{2}+\frac{\exp{(-\Gamma^{2} \nu^{2}/2)}}{\sqrt{2\pi}\Gamma\nu}),
\ee
where $\Gamma^2 = \gamma^2 / (1-\gamma^2)$ as $\gamma^2 = \langle \delta \delta' \rangle ^2 / (\langle \delta^2 \rangle \langle \delta^{'2} \rangle )$ (prime indicates the derivative with respect to the smoothing scale). Then the bias will be
\be
b^{h}_{\text{up},10}=\frac{\nu^{2}-1}{\delta_{c}}+\frac{1}{f_{\text{up}}(s)}\frac{\exp{(-\frac{\nu^{2}{(1+\Gamma)}^{2}}{2})}}{4\pi\Gamma s^{3/2}\nu}.
\ee
In the case of modulated power spectrum, the only change in the bias parameter is introduced via the redefinition of the variance and other multipoles of power spectrum in terms of mass. Otherwise, the functionality of the bias parameter remains the same.
In the case the bias parameter has a k-dependence due to non-Gaussianity or non-Markovianity, we express it in the form
\be\label{eq:kbias}
\tilde{b}^{\text{L}}_{h}(k)= b^{\text{L}}_{h}+ g(k) b^{\text{L}}_{h},
\ee
where $\tilde{b}^{\text{L}}_{h}(k)$ and $b^{\text{L}}_{h}$ are scale dependent and independent halo bias respectively and $g(k)$ is determined by non-Gaussianity or non-Markovianity extensions. Now by using equation (\ref{eq:int}) we obtain the k-dependent Lagrangian ionized fraction bias parameter
\be\label{eq:kbias2}
 \tilde{b}^{\text{L}}_{{x}_{\text{HII}}}(k)= b^{\text{L}}_{{x}_{\text{HII}}}+ g(k) b^{\text{L}}_{{x}_{\text{HII}}}.
 \ee
The above equation is used  in the main text to find the angular power spectrum of 21-cm brightness temperature in case of a local type non-Gaussianity.

\section{cosmic variance and SKA characteristics}\label{app2}

In this appendix, we will discuss the specifications of SKA observatory {as one of the main future cosmological probes} for studying the distribution of neutral Hydrogen in higher redshifts. As mentioned before, we study the effect of both long mode modulation and non-Gaussianity of local type on the angular power spectrum of 21-cm brightness which can be an observational probe for the physics beyond standard model. A very crucial point to indicate here again is that the LM and PNG-local type signal become important in large angular scales (low multipoles).
SKA as a large scale survey of 21-cm signal is a promising mission for this type of studies. SKA1 as the first phase of the project which is partially designed will be the focus of our observational prospect.
SKA1 has different sub-categories. The one which is in our interested frequency range, is SKA1-LOW telescope array. This array contains nearly 130,000 antenna elements sensitive to frequencies from 50 to 350 MHz \cite{Dewdney:2015} \footnote{https://www.skatelescope.org/key-documents/}. The detectability of the LM-PNG signals with future observations needs a very thorough and detailed analysis which is not in the scope of this work. However, in order to give an approximate touch to the problem we will discuss the main systematic noises of the SKA observation in comparison to the main signal as it comes below. The systematic errors on angular power spectrum has two main contributions: a) cosmic variance, b) thermal noise of an interferometer.
At low multipoles the {main source of uncertainty} will be the cosmic variance, since we have a finite number of measurements of the angular power spectrum: \cite{Campbell:2015mpa}
\be
\Delta C_l^{\text{cv}}=\sqrt{\frac{2}{2l+1}}C_{l}.
\ee
{Now we can compare the amplitude of this uncertainty with LM-PNG signals (in each multipole) as an approximation on the detectability of these signals.  In Figs. \ref{fig:dipole_diffrent_redshifts_new}, \ref{fig:ng_dipole_diffrent_redshifts_new}, \ref{fig:z10_ng_effect_on_lm_plot}, \ref{fig:z10_final_plot} the upper $x$-axis shows $\sqrt{\frac{2}{2l+1}}$ as the ratio of cosmic variance to the $C_{l}$. In Figs. \ref{fig:NG_z10_new}, \ref{fig:NG_diffrent_redshifts_1_new} shaded regions show the cosmic variance uncertainty area ( in Fig. \ref{fig:NG_z10_new}, the shaded region is related to $C_{l}\pm\Delta C_l^{\text{cv}}=(1\pm \sqrt{\frac{2}{2l+1}})C_{l}$  and  in Fig. \ref{fig:NG_diffrent_redshifts_1_new}) we plot  $\pm\Delta C_l^{\text{cv}}/C_{l}=\pm \sqrt{\frac{2}{2l+1}}$. This comparison is essential in a sense that the signals which are in the order of the cosmic variance can not be detected in future observations. }
{The other source of uncertainty} is the thermal noise of the interferometer which can be given by
\be
C_l^N=\frac{{(2\pi)}^3T_{\text{sys}}^{2}}{\Delta \nu t_{\text{obs}}f_{\text{cover}}^{2}l_{\text{max}}^{2}},
\ee
where $T_{\text{sys}}$ is the system temperature, $\Delta \nu$ is the bandwidth of the survey, $t_{\text{obs}}$ is the total observation time, $f_{\text{cover}}$ is the coverage fraction of the survey (effective collecting area of the core array $A_{e}$ divided by $\pi {(D_{\text{base}}/2)}^2$ where $D_{\text{base}}$ is the maximum baseline of the core array), and $l_{\text{max}}=2\pi D_{\text{base}}/\lambda$ is the highest multipole observable of the array at wavelength $\lambda$ \cite{Pritchard:2015fia}. Note that as we mentioned earlier at low multipoles the {dominant source of uncertainty} is cosmic variance and thermal noise will be important at high multipoles \cite{Shiraishi:2016omb}. The system temperature at high redshifts (for SKA1-LOW) is given by $T_{\text{sys}}= T_{\text{rec}}+ T_{\text{gal}}$ where $T_{\text{rec}}=0.1T_{\text{gal}}+40\text{K}$ is the receiver noise temperature and the contribution of galactic synchrotron radiation is given approximately by $T_{\text{gal}}= 60\times{(\nu/300 \text{MHz})}^{-2.55}\text{K}$ \cite{Dewdney:2013}. Also as mentioned by \cite{Dewdney:2015} at frequencies above $110MHz$ the ratio of $(A_{e}/T_{\text{sys}})$ is almost flat . At $z=10$ ($\nu\sim130\text{MHz}$) we set  $D_{\text{base}}=1\text{km}$ \cite{Bacon:2018dui} and $(A_{e}/T_{\text{sys}})\sim 559 \text{m}^2/\text{K}$ \cite{Dewdney:2015} so
\begin{equation}
\begin{aligned}
C_l^N&=\frac{{(2\pi)}^3(\pi (D_{\text{base}}/2)^2)^{2}}{\Delta \nu t_{\text{obs}}l_{\text{max}}^{2}}\times (\frac{A_{e}}{T_{\text{sys}}})^{-2}\\&\sim (2.3\times 10^{-12}\text{K}^2)(\frac{1000\text{hr}}{t_{\text{obs}}})(\frac{8\text{MHz}}{\Delta \nu}),
\end{aligned}
\end{equation}
or by using equation(8)
\be
\frac{C_l^N}{{(\bar{\delta T_b})}^{2}}\sim (2.7\times10^{-9})(\frac{1000\text{hr}}{t_{\text{obs}}})(\frac{\text{8MHz}}{\Delta \nu}).
\ee
 This means that the thermal noise is lower than cosmic variance at low multipoles and is comparable to our weakest signal ($\Delta C_{l}^{TT,\text{LM}}$) at $z=10$ (Fig. \ref{fig:dipole_z10_new}). \\

%

\bsp	
\label{lastpage}
\end{document}